\documentclass[prd,aps,showpacs,nofootinbib,preprint,tightenlines]{revtex4}
\usepackage{epsfig}
\usepackage[latin1]{inputenc}
\usepackage{float,amsmath,slashed}
\usepackage{graphicx}

\newcommand{\beq}{\begin{equation}}
\newcommand{\eeq}{\end{equation}}
\newcommand{\bqa}{\begin{eqnarray}}
\newcommand{\eqa}{\end{eqnarray}}

\def\lsim{\mathrel{\rlap{\lower4pt\hbox{\hskip1pt$\sim$}}
    \raise1pt\hbox{$<$}}}                % less than or approx. symbol
\def\gsim{\mathrel{\rlap{\lower4pt\hbox{\hskip1pt$\sim$}}
    \raise1pt\hbox{$>$}}}                % greater than or approx. symbol
    
\begin{document}

\title{Fermionic Collective Modes of an Anisotropic Quark-Gluon Plasma}

\author{Bj\"orn Schenke}
\affiliation{Institut f\"ur Theoretische Physik \\
  Johann Wolfgang Goethe - Universit\"at Frankfurt \\
  Max-von-Laue-Stra\ss{}e~1,
  D-60438 Frankfurt am Main, Germany\vspace*{5mm}}
\author{Michael Strickland}
\affiliation{Frankfurt Institute for Advanced Studies \\
  Johann Wolfgang Goethe - Universit\"at Frankfurt \\
  Max-von-Laue-Stra\ss{}e~1,
  D-60438 Frankfurt am Main, Germany \\}

\begin{abstract}
We determine the fermionic collective modes of a quark-gluon 
plasma which is anisotropic in momentum space.  We calculate 
the fermion self-energy in both the imaginary- and real-time 
formalisms and find that numerically and analytically (for two
special cases) there are no unstable fermionic modes.
In addition we demonstrate that in the hard-loop 
limit the Kubo-Martin-Schwinger condition, which relates 
the off-diagonal components of the real-time fermion self-energy, holds 
even for the anisotropic, and therefore non-equilibrium, quark-gluon plasma
considered here.  
The results obtained here set the stage for
the calculation of the non-equilibrium photon production rate
from an anisotropic quark-gluon plasma.
\end{abstract}
\pacs{11.15Bt, 04.25.Nx, 11.10Wx, 12.38Mh}
\maketitle
\newpage

\small

\section{Introduction}

The ultrarelativistic heavy ion collision experiments 
ongoing at the Brookhaven Relativistic Heavy Ion Collider 
(RHIC) and planned at the CERN Large Hadron Collider (LHC) 
will study the behavior of nuclear matter under extreme 
conditions.  Specifically, these experiments will explore 
the QCD phase diagram at large temperatures and small quark 
chemical potentials.  Based on the data currently available 
from the RHIC collisions there is some evidence that an 
isotropic and thermalized state has been created at times on 
the order of 1 fm/c~\cite{Gyulassy:2004zy,Gyulassy:2004vg,Heinz:2002un,Heinz:2002gs}.
The fact that thermalization proceeds rather 
rapidly is in contradiction with estimates from leading order 
equilibrium perturbation theory. However, to truly 
understand how the plasma evolves and thermalizes one has to 
go beyond the equilibrium description and study the dynamics 
of a non-equilibrium quark-gluon plasma.  In addition, it is
important to know how any deviations from equilibrium affect observables
so that one might be able to gauge how close the system truly
is to being isotropic and thermal.

For example, one would like to know how a momentum-space 
anisotropy in the distribution function of the hard modes 
would affect observables which are sensitive to the 
earliest times of quark-gluon plasma evolution when the 
anisotropy is expected to be largest.  The best signatures 
in this regard are electromagnetic probes such as photon and 
dilepton production since these particles escape the plasma 
without strong final state interactions.  In order to 
calculate in-medium photon production, however, it is 
necessary to include the effects of medium-induced fermion 
masses which serve to screen infrared divergences in the 
calculation of production cross sections.  In equilibrium 
this can be done self-consistently within the hard thermal 
loop framework \cite{Pisarski:1988vd,Braaten:1989mz,Pisarski:1990ds} 
and there are now many papers dedicated to the calculation of
equilibrium photon production at leading and next-to-leading order  
\cite{Shuryak:1978ij,Kajantie:1981wg,Halzen:1981kz,Kajantie:1982nj,Sinha:1983jm,
Hwa:1985xg,Staadt:1985uc,Neubert:1989hu,Kapusta:1991qp,Baier:1991em,Baier:1993zb,
Aurenche:1998nw,Aurenche:1999tq,Aurenche:2000gf,Steffen:2001pv,Peitzmann:2001mz,
Arnold:2001ba,Arnold:2001ms,Peitzmann:2002pz,Arnold:2002zm}.  In addition, there 
have been calculations of electromagnetic signatures from a 
plasma which is not chemically equilibrated 
\cite{Shuryak:1992bt,Dumitru:1993us,Strickland:1994rf,Traxler:1994hy,Traxler:1995kx,Kampfer:1994rr,Srivastava:1996qd,Baier:1996if,Baier:1997xc}.  
However, the problem of photon and dilepton production from a quark-gluon
plasma which is not isotropic in momentum space has not 
yet been considered.  Here we set the stage for such a 
calculation by computing the quark self-energy in such an
anisotropic plasma.

Momentum-space anisotropic distribution functions are 
relevant because of the rapid longitudinal expansion of the
partonic matter created in a heavy ion collision.  This rapid
longitudinal expansion implies that at proper times $\tau > \langle p_T 
\rangle^{-1}$, where $p_T$ is the typical transverse 
partonic momentum of the nuclear wavefunction, the parton 
distribution functions are oblate in momentum space with 
$\langle p_T \rangle > \langle p_L \rangle$. For RHIC energies
this implies that the distribution is oblate for $\tau \gsim$
0.2 fm/c and for LHC $\tau \gsim$ 0.1 fm/c.  Such an 
anisotropic quark-gluon plasma is qualitatively different 
from an isotropic one since the gluonic collective 
modes can then be unstable 
\cite{Mrowczynski:1993qm,Mrowczynski:1994xv,Mrowczynski:1996vh,Mrowczynski:2000ed,
Birse:2003qp,Randrup:2003cw,Romatschke:2003ms,Arnold:2003rq,Romatschke:2004jh,
Mrowczynski:2004kv,Dumitru:2005gp,Manuel:2005mx,Arnold:2005vb,Rebhan:2005re,Romatschke:2005pm,
Dumitru:2006pz,Schenke:2006xu,Romatschke:2006nk,Romatschke:2006wg} . The presence of these gluonic instabilities can 
dramatically influence the system's evolution leading, in 
particular, to its faster isotropization and equilibration. 
Treating this problem in all of its generality is a daunting 
task.  In order to make analytic progress we consider here the limit of 
high particle momentum scale (large $p_T$) and small coupling in order
to calculate the fermionic self-energy
in the hard-loop approximation.

In two previous papers by Paul Romatschke and one of us 
\cite{Romatschke:2003ms,Romatschke:2004jh}, we calculated 
the hard-loop gluon polarization tensor in the case that the 
momentum space anisotropy is obtained from an isotropic 
distribution by the rescaling of one direction in momentum 
space (corresponding to stretching or squeezing of the particle
distribution function along a 
special direction in momentum space). In this paper we 
extend this exploration of the collective modes of an 
anisotropic quark-gluon plasma by studying the quark 
collective modes using the same framework. Specifically, we 
derive integral expressions for the quark self-energy for 
arbitrary anisotropy and evaluate these numerically using 
the momentum-space rescaling used in the previous papers. 
We show for quarks there are still only two stable 
quasiparticle modes and no unstable modes using the momentum-space
rescaled distribution functions.  The result is similar to
the case of the fermionic collective modes in a two-stream
system \cite{Mrowczynski:2001az} where it was also found that
there were no unstable modes.\footnote{Note 
that if there were, in fact, fermionic unstable modes one 
would expect extra generation of fermions and anti-fermions 
which would naively increase electromagnetic emission from 
the plasma.}

The absence of unstable fermionic modes is expected on 
physical grounds due to the fact that fermion exclusion 
precludes the condensation of modes; however, it could be 
possible that, through pairing, fermions could circumvent 
this as has been predicted \cite{fp1a,fp1b,fp2a,fp2b,fp2c} 
and demonstrated \cite{fp3} in superfluid condensation of 
cold fermionic atoms. However, this would require a description in 
terms of fermionic bound or composite states which are not 
included at the level of hard loops so we do not expect to 
find any fermionic condensate-like instabilities using this 
approximation.
This is verified via an explicit contour 
integration of the inverse hard-loop quark propagator for 
the two special cases in which we can obtain analytic 
expressions for the self-energy. The special cases 
considered analytically are (a) the case when the wave 
vector of the collective mode is parallel to the anisotropy 
direction with arbitrary oblate anisotropy and (b) for all 
angles of propagation in the limit of an infinitely oblate 
anisotropy.

Finally, we present a calculation of the off-diagonal 
components of the anisotropic fermion self-energy using the 
real-time formalism of quantum field theory.  Using this 
explicit calculation we demonstrate that within the hard-loop 
framework the high-temperature limit of the Kubo-Martin-Schwinger 
(KMS) formula, namely $\Sigma_{12} = - \Sigma_{21}$, holds even for the non-equilibrium 
configuration considered here.  This is a non-trivial result 
since relations of this kind can only be proven to hold in 
an equilibrated plasma.  If generic, this implies that a 
kind of generalized KMS condition applies also in a non-equilibrium setting.

The organization of the paper is as follows: In Section
\ref{quarkse} we derive integral expressions for the retarded quark self-energy 
in a system with an anisotropic distribution obtained from
contracting an isotropic distribution in one direction. We show
plots of the different components of this self-energy for
different anisotropy strengths and various orientations of the
wave vector with respect to the direction of the anisotropy. We
point out the strong dependence of the self-energy on the strength
of the anisotropy and the angle of propagation with respect to the
anistropy direction. In Section \ref{sec:nyquist} we prove
analytically that for the case that the wave vector of the
collective mode lies in the direction of the anisotropy there are
no unstable modes. The same proof is performed in Section
\ref{sec:largexi} for the extremely anisotropic limit and
arbitrary orientation of the wave vector. In Section \ref{sec:realtime}
we extend our previous results to the real-time formalism and
compare with the results obtained in the imaginary time formalism.

\section{Anisotropic quark self-energy}
\label{quarkse}

The integral expression for the retarded quark self-energy for an anisotropic system has been
obtained previously \cite{Mrowczynski:2000ed} and is given by
\bqa \label{q-self} \Sigma(K) = {C_F \over 4} g^2 \int_{\bf p} {
f ({\bf p}) \over |{\bf p}|} {P \cdot \gamma \over P\cdot
K} \;,\label{retself}
\eqa
where $C_F \equiv (N_c^2 -1)/2N_c$, $\int_{\bf p} = \int d^3 p/(2 \pi)^3 $, and
$$
f ({\bf p}) \equiv 2 \left( n({\bf p}) + \bar n ({\bf p})
\right) + 4 n_g({\bf p}) \; .
$$
To simplify the calculation we follow Ref.~\cite{Romatschke:2003ms} and require the
distribution function $f({\bf p})$ to be given by
\beq
f({\bf p})=f_{\xi}({\bf p}) = N(\xi) \ %
f_{\rm iso}\left(\sqrt{{\bf p}^2+\xi({\bf p}\cdot{\bf \hat n})^2}\right)%
\; .
\label{squashing}
\eeq
Here ${\bf \hat n}$ is the direction of the anisotropy, $\xi>-1$ is a
parameter reflecting the strength of the anisotropy and $N(\xi)$ is a
normalization constant.  For the application to heavy ion collisions
${\bf \hat n}$ is the beamline (longitudinal) direction and the relevant
anisotropy parameter at times $\tau > \langle p_T \rangle^{-1}$ is
positive, $\xi>0$, corresponding to an oblate distribution.

To fix $N(\xi)$ we require that the number density
to be the same both for isotropic and arbitrary anisotropic systems,
\beq
\int_{\bf p} f_{\rm iso}(p)=%
\int_{\bf p} f_{\xi}({\bf p})=N(\xi)%
\int_{\bf p} %
f_{\rm iso}\left(\sqrt{{\bf p}^2+\xi({\bf p}\cdot{\bf \hat n})^2}\right) ,
\eeq
and can be evaluated to be
\beq
N(\xi)=\sqrt{1+\xi}.
\eeq
Using Eq.~(\ref{squashing}) and performing the change of variables
\begin{equation}
\tilde{p}^2=p^2\left(1+\xi ({\bf v}\cdot{\bf \hat{n}})^2\right) \; ,
\label{variablechange}
\end{equation}
we obtain
\bqa \label{q-self2}
\Sigma(K) = m_q^2 \sqrt{1+\xi}
\int {d\Omega\over4\pi} \left(1+\xi (\hat{\bf p}\cdot\hat{\bf n})^2 \right)^{-1}
{P \cdot \gamma \over P\cdot K} \;,
\eqa
where
\bqa
m_q^2 = {g^2 C_F \over 8 \pi^2} \int_0^\infty dp \,
    p \, f_{\rm iso}(p) \; .
\eqa
We then decompose the self-energy into four contributions
\bqa \Sigma(K) = \gamma^0 \Sigma_0 + {\boldsymbol\gamma}\cdot{\mathbf
\Sigma}\; .
\eqa

  \begin{figure}[htb]
    \begin{center}
        \includegraphics[height=5cm]{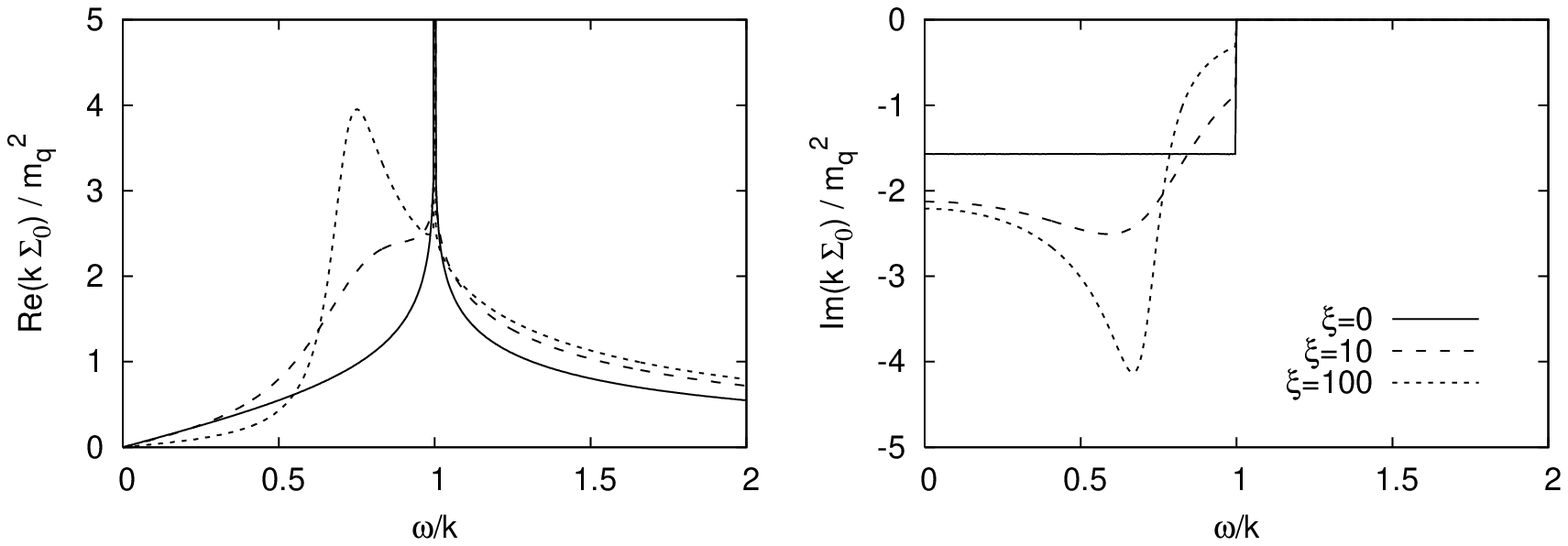}
        \caption{Real and imaginary part of $\Sigma_0$ as a function of $\omega/k$ for $\theta_n=\pi/4$ and $\xi=\{0,10,100\}$.}
        \label{fig:sigma0}
        \vspace{.3in}
        \includegraphics[height=5cm]{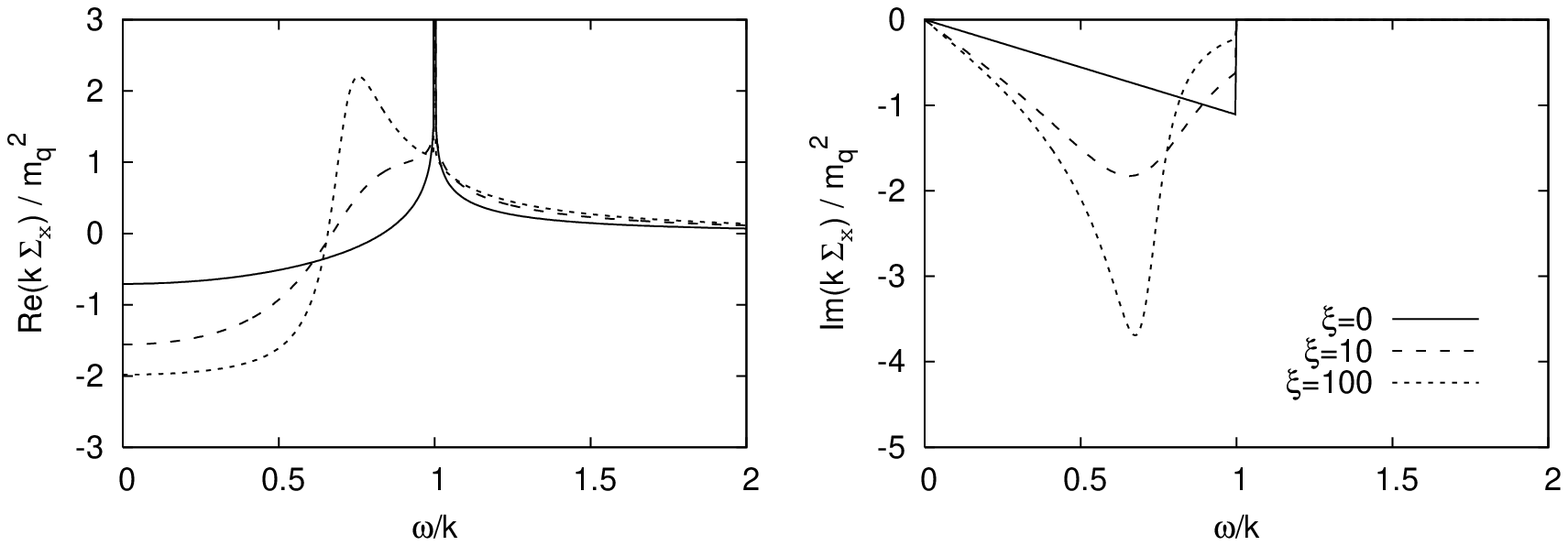}
        \caption{Real and imaginary part of $\Sigma_x$ as a function of $\omega/k$ for $\theta_n=\pi/4$ and $\xi=\{0,10,100\}$.}
        \label{fig:sigma1}
        \vspace{.3in}
        \includegraphics[height=5cm]{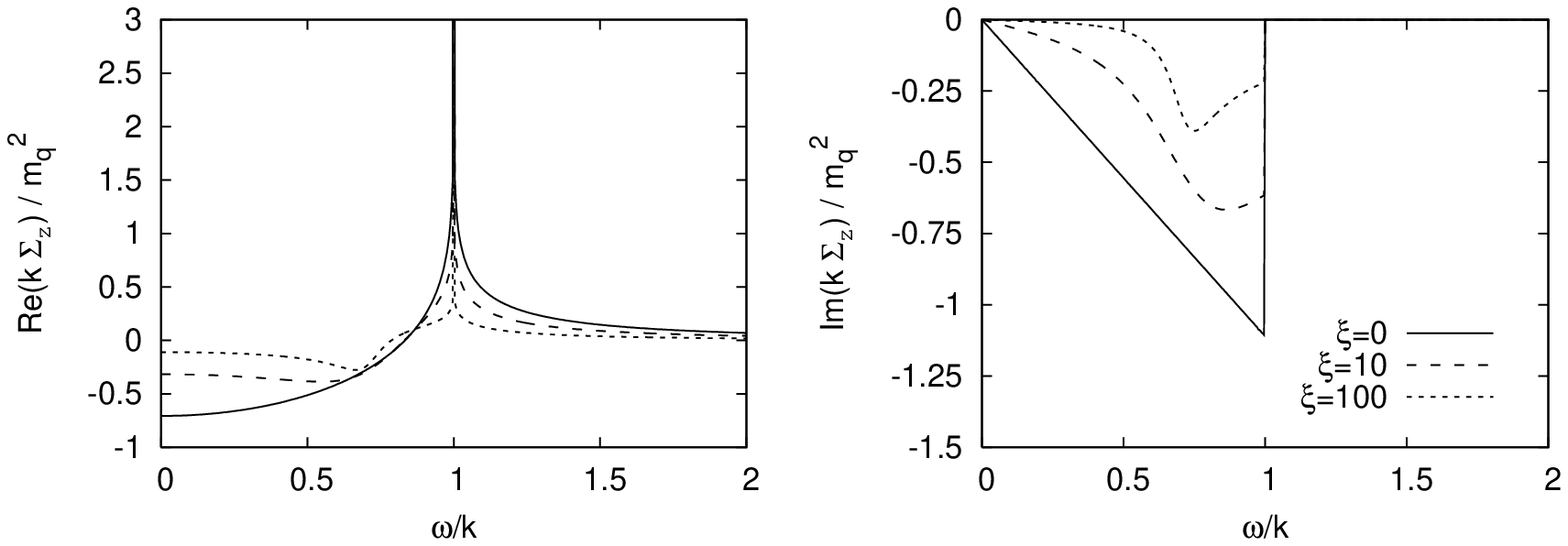}
        \caption{Real and imaginary part of $\Sigma_z$ as a function of $\omega/k$ for $\theta_n=\pi/4$ and $\xi=\{0,10,100\}$.}
        \label{fig:sigma2}
    \end{center}
  \end{figure}

The fermionic collective modes are determined by finding all
four-momenta for which the determinate of the inverse propagator
vanishes
\bqa
{\rm det}\,S^{-1} = 0 \; ,
\eqa
where
\bqa
i S^{-1}(P) &=& \gamma^\mu p_\mu - \Sigma \, , \nonumber \\
&\equiv& \gamma^\mu A_\mu \;.
\eqa
with $A(K)=(k_0 - \Sigma_0,{\bf k} - {\bf \Sigma})$.
Using the fact that ${\rm det}(\gamma^\mu A_\mu) = (A^\mu A_\mu)^2$
and defining $A_s^2 = {\bf A}\cdot{\bf A}$ we obtain
\bqa
A_0 = \pm A_s \, .
\label{fermiondisp}
\eqa

In practice, we can define the $z$-axis to be in the $\hat{\bf n}$
direction and use the azimuthal symmetry to restrict our consideration
to the $x\!-\!z$ plane.  In this case we need only three functions
instead of four
\bqa
\Sigma_0(w,k,\theta_n,\xi) &=& {1\over2} m_q^2 \sqrt{1+\xi} \int_{-1}^1 dx
    { R(w-k\cos\theta_n x,k\sin\theta_n\sqrt{1-x^2})
     \over 1 + \xi x^2 } \; , \nonumber \\
\Sigma_x(w,k,\theta_n,\xi) &=& {1\over2} m_q^2 \sqrt{1+\xi}
\int_{-1}^1 dx
    { \sqrt{1-x^2} S(w-k\cos\theta_n x,k\sin\theta_n\sqrt{1-x^2})
     \over 1 + \xi x^2 } \; , \nonumber \\
\Sigma_z(w,k,\theta_n,\xi) &=& {1\over2} m_q^2 \sqrt{1+\xi}
\int_{-1}^1 dx
    { x R(w-k\cos\theta_n x,k\sin\theta_n\sqrt{1-x^2})
     \over 1 + \xi x^2 } \; ,\label{selfenergies}
\eqa
where
\bqa
R(a,b) &=& \left(\sqrt{a+b+i\epsilon}\sqrt{a-b+i\epsilon}\right)^{-1} \, , \nonumber \\
S(a,b) &=& {1\over b}\left[a R(a,b) - 1\right] \, . \eqa

\begin{figure}[t]
    \begin{center}
        \includegraphics[height=5.25cm]{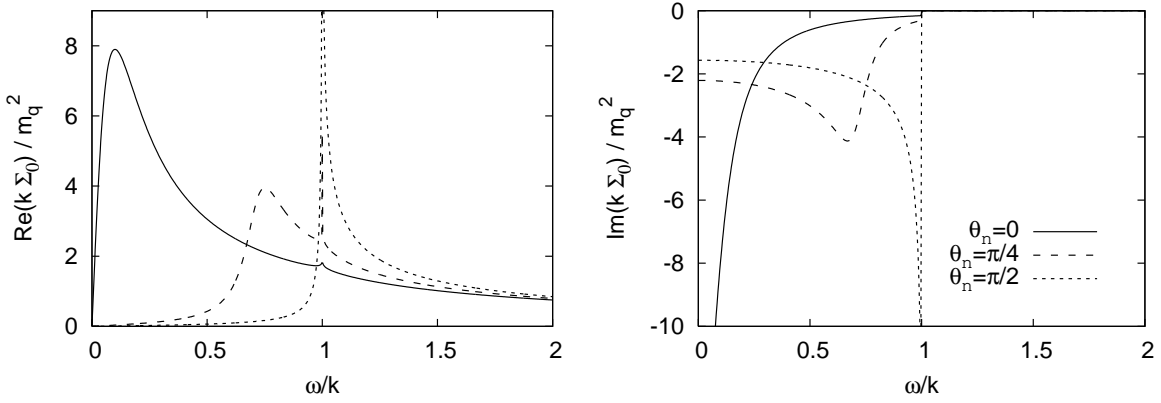}
        \caption{Real and imaginary part of $\Sigma_0$ as a function of $\omega/k$ for $\xi=100$ and $\theta_n=\{0,\pi/4,\pi/2\}$.}
        \label{fig:sigmaxi100}
    \end{center}
  \end{figure}

In Figs.~\ref{fig:sigma0} through \ref{fig:sigma2} we plot the
real and imaginary parts of the quark self-energies $\Sigma_0$,
$\Sigma_x$, and $\Sigma_z$ for $\xi=\{0,10,100\}$. From these
Figures we see that the spacelike quark self-energy is strongly
affected by the presence of an anisotropy with a peak appearing at
$\omega/k = \sin\theta_n$ for strong anisotropies. To further
illustrate this in Fig.~\ref{fig:sigmaxi100} we have plotted
$\Sigma_0$ for $\xi=100$ and $\theta_n=\{0,\pi/4,\pi/2\}$. From
this Figure we see that there is a large directional dependence of
the spacelike quark self-energy.  Note that this could have a
measurable impact on quark-gluon plasma photon production during
the early stages of evolution since screening of infrared
divergences in leading order photon production amplitudes requires
as input the hard-loop fermion propagator for spacelike momentum.
We return to this point in Section \ref{sec:realtime} and sketch how to
calculate photon emission from an anisotropic quark-gluon plasma.
Assuming the necessary measurements of the rapidity dependence of the thermal
photon spectrum could be performed, photon emission could provide an
excellent measure of the degree of momentum-space anisotropy in
the partonic distribution functions at early stages of a heavy-ion
collision.

For general $\xi$ and $\theta_n$ we have to evaluate
the integrals given in Eq.~(\ref{selfenergies}) numerically.
To find the collective modes we then numerically solve the 
fermionic dispersion relations given by Eq.~(\ref{fermiondisp}).
As in the isotropic case, for real timelike momenta 
($\mid\!\!\omega\!\!\mid>\mid\!\!k\!\!\mid$, ${\rm Im}(\omega/k)=0$) there are two stable 
quasiparticle modes which result from choosing either plus 
or minus in Eq.~(\ref{fermiondisp}).\footnote{Note that there are four solutions 
to the dispersion relations since each solution exists at both positive and negative $\omega$.}  
We have looked for 
modes in the upper- and lower-half planes and numerically we 
find none.  In the next section we explicitly count the 
number of modes using complex  contour integration and 
demonstrate that there are no unstable collective modes in 
two special cases.

\subsection{Special case: $\mathbf{k}\parallel\mathbf{\hat{n}}$}
\label{sec:nyquist} Let us consider the special case where the
momentum of the collective mode is in the direction of the
anisotropy $\mathbf{k}\parallel\mathbf{\hat{n}}$, i.e.,
$\theta_n=0$. In this case the integrals in Eq.
(\ref{selfenergies}) can be evaluated analytically. $\Sigma_x$
becomes zero, while the other components read
\begin{align}
    \Sigma_0(\omega,k,\theta_n=0,\xi)&=\frac{1}{2}m_q^{2}\frac{\sqrt{1+\xi}}{\xi\omega^2+k^2}\left[2\sqrt{\xi}\omega\arctan\sqrt{\xi}+k\ln\left(\frac{\omega+k}{\omega-k}\right)\right]\notag\\
    \Sigma_z(\omega,k,\theta_n=0,\xi)&=\frac{1}{2}m_q^{2}\frac{\sqrt{1+\xi}}{\xi\omega^2+k^2}\left[-2\frac{1}{\sqrt{\xi}}k\arctan\sqrt{\xi}+\omega\ln\left(\frac{\omega+k}{\omega-k}\right)\right]\notag\,\text{.}\\
\end{align}
Eq. (\ref{fermiondisp}) simplifies to
\begin{equation}
    \omega-\Sigma_0=\pm(k-\Sigma_z)
    \label{fermiondispspecial}\,\text{.}
\end{equation}

\subsubsection*{Nyquist analysis}

\begin{figure}[t]
    \begin{center}
        \includegraphics[height=6cm]{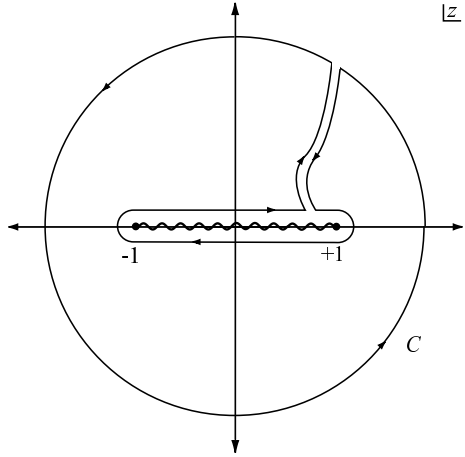}
        \caption{Contour $C$ in the complex $z$-plane used for the Nyquist analysis.}
        \label{fig:nyquistcontour}
    \end{center}
  \end{figure}

We now show analytically for this special case that unstable modes
do not exist. This is done by a Nyquist analysis of the following
function:
\begin{equation}
    f_{\mp}(\omega,k,\xi)=\omega-\Sigma_0(\omega,k,\xi)\mp\left[k-\Sigma_z(\omega,k,\xi)\right]\label{function}\,\text{.}
\end{equation}
In practice, that means that we evaluate the contour integral
\begin{equation}
    \frac{1}{2\pi i}\oint_C dz\frac{f^{\prime}_{\mp}(z)}{f_{\mp}(z)}=N-P\,\text{,}\label{nyquist}
\end{equation}
which gives the numbers of zeros $N$ minus the number of poles $P$
of $f_{\mp}$ in the region encircled by the closed path $C$. In
Eq. (\ref{nyquist}) and in the following, we write the functions
$f_{\mp}$ in terms of $z=\omega/k$ and for clarity do not always
state the explicit dependence of $f_{\mp}$ on $k$ and $\xi$.
Choosing the path depicted in Fig. \ref{fig:nyquistcontour}, which
excludes the logarithmic cut for real $z$ with $z^2<1$ of the
function (\ref{function}), leads to $P=0$ and the left hand side
of Eq. (\ref{nyquist}) equals the number of modes $N$. Evaluation
of the respective pieces of the contour $C$ for each $f_{-}$ and
$f_{+}$ leads to
\begin{equation}
    N_{\mp}=1+0+0+1=2\,\text{,}\label{number}
\end{equation}
such that for the total number we get is $N=N_{-}+N_{+}=4$, which
corresponds to the stable modes (two for
positive $\omega$ and two for negative $\omega$). The four contributions in
(\ref{number}) are the following:
\begin{enumerate}
    \item The first $1$ results from integration along the large circle at $|z|\gg 1$.
    \item The first zero is the contribution from the path connecting
          the large circle with the contour around $z=\pm 1$.
    \item The second zero stems from the two small half-circles around $z=\pm 1$
    \item The last $1$ is obtained from integration along the straight lines running
          infinitesimally above and below the cut between $z=-1$ and
          $z=1$. See below for details on this integration.
\end{enumerate}

The last contribution can be evaluated using
\begin{equation}
    \int_{-1 + i\epsilon}^{1 + i\epsilon}
    dz\frac{f^{\prime}_{\mp}(z)}{f_{\mp}(z)}=\ln\frac{f_{\mp}(1 + i\epsilon)}{f_{\mp}(-1 + i\epsilon)}+2\pi i n\,\text{,}\label{straightlines}
\end{equation}
for the line above and the corresponding expression for the line
below the cut. $n$ is the number of times the function $f_{\mp}$
crosses the logarithmic cut located on the real axis, running from
zero to minus infinity. This cut is due to the appearance of the
logarithm on the right hand side of Eq. (\ref{straightlines}). In
the sum of the line integrations above and below the cut diverging
contributions from the first part on the right hand side of Eq.
(\ref{straightlines}) cancel and we are left with a contribution
of $2\pi i$ for each function. Furthermore it is necessary to show
that neither $f_{-}$ nor $f_{+}$ crosses the cut. The proof is
given in some detail for $f_{-}$ and is performed analogously for
$f_{+}$. From Eq. (\ref{function}) we find for $f_{-}$:
\begin{align}
    f_{-}(z,k,\xi)&=z-1+\frac{\sqrt{1+\xi}}{2(1+\xi
    x^2)}\frac{1}{k^2}\left[-2\left(z+\frac{1}{\xi}\right)\sqrt{\xi}\arctan\sqrt{\xi}+(z-1)\ln\left(\frac{z+1}{z-1}\right)\right]\,\text{.}
\end{align}
We want to study whether this function crosses the real axis in
the range $\text{Re}[z]\in[-1,1]$ for $\text{Im}[z]\rightarrow0$,
i.e., whether the imaginary part of $f_{-}$ changes sign in that
range. On the straight line infinitesimally above the cut the
imaginary part of $f_{-}$ is given by
\begin{equation}
    \text{Im}\left[
    \lim_{\epsilon\rightarrow0}f_{-}(x+i\epsilon,k,\xi)\right]=-\frac{\pi}{2}\frac{\sqrt{1+\xi}(x-1)}{k^2(1+\xi
    x^2)}\,\text{,}\label{impartf}
\end{equation}
for real $x$. It is only zero for $x=1$, which means that the
function $f_-$ can not cross but merely touch the cut within the
limits of the integration. On the straight line below the cut we
get the same result (\ref{impartf}) with a minus sign. For
$f_{+}$, we find that the imaginary part in the regarded range
only becomes zero for $x=-1$, which means that the logarithmic cut
is not crossed within $[-1,1]$ either. Hence we have proved for
the case $\mathbf{k}\parallel\mathbf{\hat{n}}$ that there are no
more solutions than the four stable modes. In
particular we have shown that unstable fermionic modes can not
exist.

\subsection{Large-$\xi$ limit}
\label{sec:largexi}
 In the extremely anisotropic case where
$\xi\rightarrow \infty$ the self-energies for arbitrary angle
$\theta_n$ can be calculated explicitly. The distribution function
(\ref{squashing}) becomes~\cite{Romatschke:2003yc}
\begin{equation}
    \lim_{\xi\rightarrow\infty}
    f_{\xi}(\textbf{p})=\delta(\mathbf{\hat{p}}\cdot\mathbf{\hat{n}})\int_{-\infty}^{\infty}dx
    f_{\text{iso}}\left(p\sqrt{1+x^2}\right)\,\text{.}\label{largexif}
\end{equation}
With $\mathbf{\hat{n}}$ in the $z$-direction this implies that
$\mathbf{p}$ lies in the $x$-$y$-plane only. As in Section
\ref{quarkse}, due to azimuthal symmetry, we consider the case
where $\mathbf{k}$ lies in the $x$-$z$-plane only. Using
(\ref{largexif}) we obtain from Eqs. (\ref{selfenergies})
\begin{align}
    \Sigma_0(\omega,k,\theta_n)&=\frac{\pi}{2}m_q^2\frac{1}{\sqrt{\omega+k \sin\theta_n}\sqrt{\omega-k
    \sin\theta_n}}\,\text{,}\notag\\
    \Sigma_x(\omega,k,\theta_n)&=\frac{\pi}{2 k \sin \theta_n}m_q^2\left(\frac{\omega}{\sqrt{\omega+k \sin\theta_n}\sqrt{\omega-k
    \sin\theta_n}}-1\right)\,\text{.}\label{lxsigma}
\end{align}
Since $p_z$ is always zero, $\Sigma_z$ vanishes. Eq.
(\ref{fermiondisp}) now becomes
\begin{equation}
    \omega-\Sigma_0=\pm\sqrt{(k_x-\Sigma_x)^2+k_z^2}\,\text{.}\label{lxdisp}
\end{equation}

\subsubsection*{Nyquist analysis}

Again, we only find four stable modes and will now show
analytically that these are the only solutions in the large
$\xi$-limit for arbitrary angle $\theta_n$.
The cut resulting from the complex square roots in (\ref{lxsigma})
can be chosen to lie between $z = -\sin\theta_n$ and 
$z = \sin\theta_n$ on the real axis.
The Nyquist analysis can then be performed analogously to that in
Section \ref{sec:nyquist} with the contour in Fig.
\ref{fig:nyquistcontour} adjusted such that the inner path still
runs infinitesimally close around the cut. Using this path in the
evaluation of Eq. (\ref{nyquist}) for the functions
\begin{equation}
    f_{\mp}(\omega,k,\theta_n)=\omega-\Sigma_0\mp
    \sqrt{(k_x-\Sigma_x)^2+k_z^2}\,\text{,}\label{lxfunc}
\end{equation}
we find the number of solutions to Eq. (\ref{lxdisp}) to be
\begin{equation}
    N_{\mp}=1+0+\frac{1}{4}+\frac{1}{4}+\frac{1}{2}=2\,\text{,}\label{lxsol}
\end{equation}
so that again there are $N=N_++N_-=4$ solutions, which are the
known stable modes. The decomposition in (\ref{lxsol}) is done as
follows:
\begin{enumerate}
    \item The first contribution to $N_{\mp}$ comes from integration along the large outer
          circle at $|z|\gg 1$.
    \item The zero stems from the paths connecting the outer and the
          inner circle.
    \item The two contributions of $1/4$ result from integrations along the small
          circles around $-\sin\theta_n$ and $\sin\theta_n$.
    \item The last contribution of $1/2$ comes from integration along the straight lines running infinitesimally
          close above and below the cut. We discuss this part in further detail below.
\end{enumerate}
The last contribution can be obtained using Eq.
(\ref{straightlines}). For the evaluation of the limit
$\epsilon\rightarrow 0$ it is essential to note that the $f_{\mp}$
behave like $\ln \epsilon$ or $1/(\ln \epsilon)$ (depending on
which function is evaluated on which line) and are both negative
as $\epsilon \rightarrow 0$. This results in a contribution of $+i\pi$ for each
function and integration, because in all cases the imaginary part
of both functions can be shown to be positive in the regarded
limit. All other contributions, including the diverging parts $\pm
\ln(-\ln\epsilon)$ cancel in the sum of the results from the upper
and lower line.

Again, we need to show that the functions $f_{\mp}$ do not cross
the logarithmic cut for $z \in [-\sin\theta_n,\sin\theta_n]$,
i.e., that $n=0$ in Eq. (\ref{straightlines}). It is possible to
find an analytic expression for the imaginary part of $f_{\mp}$
using
\begin{equation}
    \text{Im}\sqrt{x+i y}=\frac{1}{\sqrt{2}}\,\text{sgn}(y)\sqrt{\sqrt{x^2+y^2}-x}\,\text{,}
\end{equation}
for the imaginary part of the square root appearing in
(\ref{lxfunc}) with real $x$ and $y$. Then the only solutions to
\begin{equation}
    \text{Im}f_{\mp}(z)=0
\end{equation}
are found analytically to be $\text{Re}(z)=\sin\theta_n$ and
$\text{Re}(z)=-\sin\theta_n$ for $f_-$ and $f_+$ respectively.
This means that the cut is not crossed during the integration
along the straight lines and that the contribution from this piece
is in fact $1/2$.

\section{Fermion self-energy from the real-time formalism}
\label{sec:realtime}

In this section we extend our previous results to the real-time 
formalism and demonstrate that the high-temperature 
limit of the Kubo-Martin-Schwinger formula, $\Sigma_{12} = -
\Sigma_{21}$, holds even for the non-equilibrium 
configuration considered here. We will use the real-time 
formulation of 
Refs.~\cite{Schw61,Ke64,Ke65,Chou:1984es,Landsman:1986uw,Carrington:1997sq}. 
In this case both propagators and self-energies become 
$2\times 2$ matrices. The free propagators are given by
\begin{align}
    S(K)=(\slashed{K}+m)&\left[\left(\begin{array}{cc}
        \frac{1}{K^2-m^2+i\epsilon} & 0 \\
        0 & \frac{-1}{K^2-m^2-i\epsilon}\end{array}
        \right)\right.\notag\\
        &~~~~~~\left.+2\pi i \delta(K^2-m^2)\left( \begin{array}{cc}
        f_{\text{F}}(K) & -\theta(-k_0)+f_{\text{F}}(K)  \\
        -\theta(k_0)+f_{\text{F}}(K) & f_{\text{F}}(K)\end{array}
        \right)\right]\text{,}\label{prop}
\end{align}
with the general fermion distribution function $f_{\text{F}}(K)$.

The components $(12)$ and $(21)$ of the self-energy matrices are
related to the emission and absorption probability of the particle
species under consideration
\cite{Chou:1984es,Mrowczynski:1992hq,Calzetta:1986cq}. To lowest
order photons are produced via annihilation and Compton processes
\begin{equation}
 q + \bar q \to g + \gamma, \qquad
 q (\bar q) + g \to q (\bar q) + \gamma \, .
\end{equation}
Within the real-time formalism the rate of photon emission can be
expressed as \cite{Baier:1997xc}
\begin{equation}
 E\frac{dR}{d^3 q} = \frac{i}{2(2\pi)^3} {\Pi_{12}}_\mu^\mu (Q) \, ,
 \label{photonrate}
\end{equation}
from the trace of the (12)-element $\Pi_{12}$ of the photon-polarization
tensor.
\begin{equation}
 -i {\Pi_{12}}_\mu^\mu(Q) = -e^2e_q^2 N_c \int \frac{d^4 p}{(2\pi)^4} Tr \left[
 \gamma^\mu \left. iS_{12}^\star(P) \right|_{HL} \gamma_{\mu} i S_{21}(P-Q) +
 \gamma^\mu iS_{12}(P) \gamma_\mu \left. iS_{21}^\star(P-Q) \right|_{HL}
 \right]\,,
\end{equation}
where $e_q$ is the quark charge.
Here $S_{12}$ and $S_{21}$ are the free fermion propagators from Eq.\,(\ref{prop}) 
and propagators with an HL subscript are the full propagators in the hard-loop
approximation.  The hard-loop propagators satisfy a fluctuation dissipation relation, which
in the quasi-static case is given by
\begin{equation}
    S_{12/21}^\star\left.(P)\right|_{HL}=\left.S_{\text{ret}}^\star(P)\right|_{HL}\Sigma_{12/21}(P)\left.S_{\text{adv}}^\star(P)\right|_{HL}\text{\,.}
\end{equation}
The retarded propagator reads
\begin{equation}
    \left.S_{\text{ret}}^\star(P)\right|_{HL}=\frac{1}{\slashed{P}-m-\Sigma(P)}\text{\,,}
\end{equation}
where $\Sigma(P)$ is the retarded self-energy given in
Eq.\,(\ref{retself}). The advanced propagator follows analogously
with the advanced self-energy and to one loop order $\Sigma_{12}$ is given by
\begin{align}
    \Sigma_{12}(P)=2 i g^2 C_\text{F} \int \frac{d^4 p}{(2\pi)^4}
    S_{12}(K)\Delta_{12}(Q)\text{\,,}
\end{align}
where $\Delta_{12}$ is the (12)-element of the matrix boson
propagator given by
\begin{align}
    \Delta(K)=\left(\begin{array}{cc}
        \frac{1}{K^2-m^2+i\epsilon} & 0 \\
        0 & \frac{-1}{K^2-m^2-i\epsilon}\end{array}
        \right)-2\pi i \delta(K^2-m^2)\left( \begin{array}{cc}
        f_{\text{B}}(K) & \theta(-k_0)+f_{\text{B}}(K)  \\
        \theta(k_0)+f_{\text{B}}(K) & f_{\text{B}}(K)\end{array}
        \right)\text{\,.}\label{propboson}
\end{align}
With the anisotropic distribution function (\ref{squashing})
$\Sigma_{12}$ can be evaluated in the hard-loop approximation to
read
\bqa
\Sigma_{12}^{\mu}(P)&=&i \frac{g^2 C_{\text{F}}}{(2\pi)^2}\int d\tilde{k}\int_{0}^{2\pi} d\phi
\int_{-1}^{+1} dx \, \frac{\tilde{k}^2}{(1+\xi x^2)^{3/2}} \nonumber \\
&& \hspace{-2cm} \times \left[\!\left.\frac{k^{\mu}}{k}\right|_{k_0=k}\!\!\delta\left(g_{-}\right)  N(\xi)f_{F}^{\text{iso}}(\tilde{k})\left( N(\xi)f_{\text{B}}^{\text{iso}}(\tilde{k})+1\right)
+\left.\frac{k^{\mu}}{k}\right|_{k_0=-k}\!\!\delta\left(g_{+}\right) N(\xi)f_{B}^{\text{iso}}(\tilde{k})\left( N(\xi)f_{\text{F}}^{\text{iso}}(\tilde{k})-1\right)\!\right]\text{\,,}
\nonumber \\
\eqa
where
\beq
g_\pm = 2\frac{\tilde{k}}{\sqrt{1+\xi
    x^2}}\left[\pm p_0+p\left(\sin\theta_n\sqrt{1-x^2}\cos\phi+\cos\theta_n
    x\right)\right]
\eeq
  \begin{figure}[t]
    \begin{center}
        \includegraphics[height=6cm]{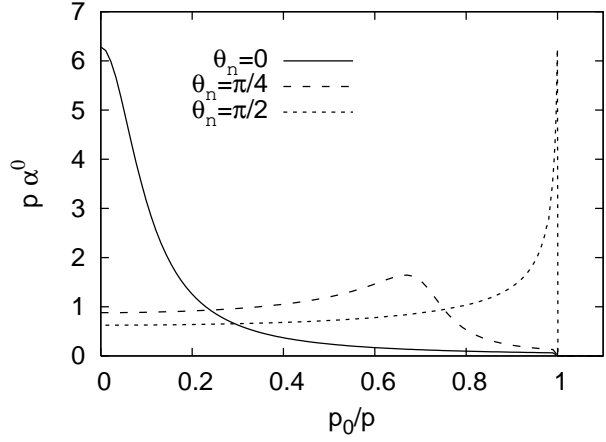}
        \caption{$\theta_n$-dependent part $\alpha^0$ of the self-energy $\Sigma_{12}^0$. $\theta_n\in\{0,\pi/4,\pi/2\}$ and $\xi=100$.}
        \label{fig:sigma12-0-xi100}
    \end{center}
  \end{figure}
and we chose $\mathbf{p}$ to lie in the $x-z$-plane and used the
change of variables (\ref{variablechange}) for $k$. Note that in the hard-loop
limit one can ignore the quark masses and hence they have been explicitly set to 
zero above.  The term
$k^{\mu}/k$ does not depend on $k$ and is given by $(\pm 1, \sin
\theta \cos \phi, \sin \theta \sin \phi, \cos \theta)$. Evaluation
of the $\delta$-function leads to
\begin{align}
    -i \Sigma_{12}^{\mu}(P,\theta_n,\xi)=&A\, \alpha^{\mu}(P,\theta_n,\xi)+B\, \beta^{\mu}(P,\theta_n,\xi)\text{\,,}\label{alphabeta}
\end{align}
with
\begin{align}
   \alpha^{\mu}(P,\theta_n,\xi)&=\int d\phi\, \sum_i \left.\frac{k^{\mu}}{k}\right|_{k_0=k}\left|\frac{(1-x_i^2)^{1/2}}{(1-x_i^2)^{1/2}(p
    \cos\theta_n+p_0\xi x_i)-p \sin\theta_n x_i(1+\xi)
    \cos\phi}\right|\theta(1-x_i^2)\notag\\
   \beta^{\mu}(P,\theta_n,\xi)&=\int d\phi\, \sum_i \left.\frac{k^{\mu}}{k}\right|_{k_0=-k}\left|\frac{(1-\tilde{x}_i^2)^{1/2}}{(1-\tilde{x}_i^2)^{1/2}(p
    \cos\theta_n-p_0\xi \tilde{x}_i)-p \sin\theta_n \tilde{x}_i(1+\xi)
    \cos\phi}\right|\theta(1-\tilde{x}_i^2)\text{\,,}
\end{align}
where the $x_i$ and $\tilde{x}_i$ are solutions to $\frac{\tilde{k}}{\sqrt{1+\xi
    x^2}}\left[-p_0+p\left(\sin\theta_n\sqrt{1-x^2}\cos\phi+\cos\theta_n
    x\right)\right]=0$ and $\frac{\tilde{k}}{\sqrt{1+\xi
    x^2}}\left[p_0+p\left(\sin\theta_n\sqrt{1-x^2}\cos\phi+\cos\theta_n
    x\right)\right]=0$, respectively, and
\begin{align}
    A&=\frac{g^2 C_{\text{F}}}{8\pi^2}\int dk\,k\,  N(\xi)f_{F}^{\text{iso}}(k)\left( N(\xi)f_{\text{B}}^{\text{iso}}(k)+1\right)\text{\,,}\\
    B&=\frac{g^2 C_{\text{F}}}{8\pi^2}\int dk\,k\,  N(\xi)f_{B}^{\text{iso}}(k)\left( N(\xi)f_{\text{F}}^{\text{iso}}(k)-1\right)\text{\,.}
\end{align}
There can be $N\in\{0,1,2\}$ solutions for both $x_i$ and $\tilde{x}_i$, depending on the parameters $p,p_0,\theta_n$ and $\phi$.
%In particular $\left[(p_0/p-x_i\cos\theta_n)\csc\theta_n\sec\phi\right]\geq0$ has to hold for $x_i$ to be a solution.
Note that $\frac{k^{\mu}}{k}$ is also given in terms of the $x_i$.
It is easily verified that
\begin{equation}
    \alpha^{\mu}(P,\theta_n,\xi)=-\beta^{\mu}(P,\theta_n,\xi)\text{\,,}
\end{equation}
  \begin{figure}[t]
    \begin{center}
        \includegraphics[height=6cm]{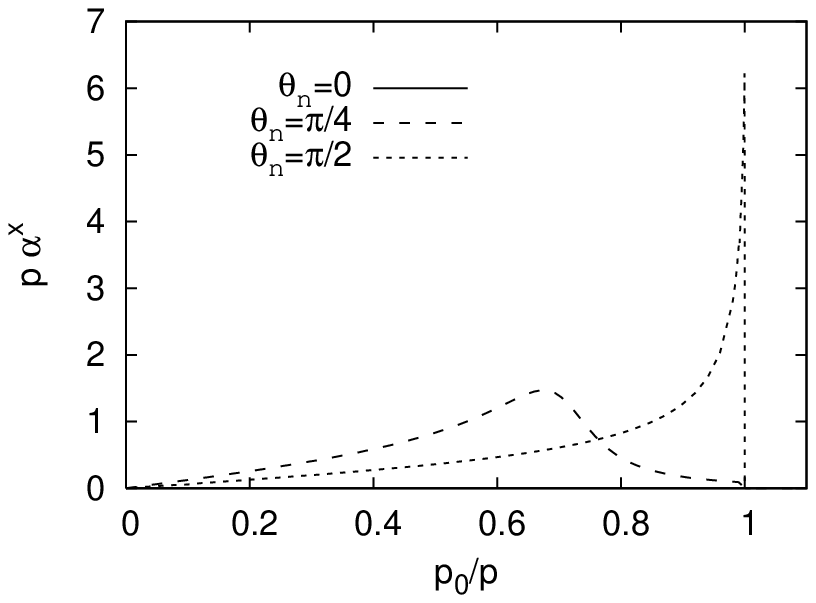}
        \caption{$\theta_n$-dependent part $\alpha^x$ of the self-energy $\Sigma_{12}^x$. $\theta_n\in\{0,\pi/4,\pi/2\}$ and $\xi=100$. For $\theta_n=0$ $\alpha^x=0$.}
        \label{fig:sigma12-x-xi100}
    \end{center}
  \end{figure}
    \begin{figure}[t]
    \begin{center}
        \includegraphics[height=6cm]{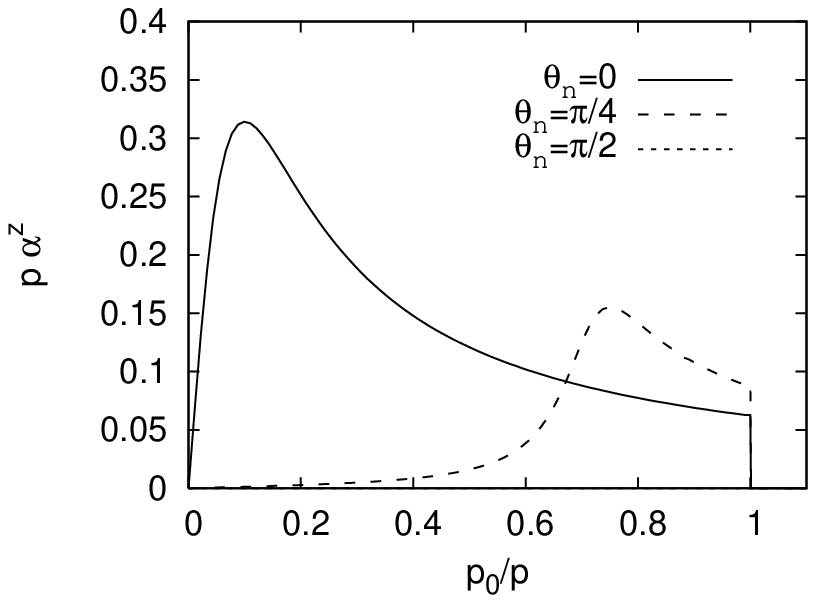}
        \caption{$\theta_n$-dependent part $\alpha^z$ of the self-energy $\Sigma_{12}^z$. $\theta_n\in\{0,\pi/4,\pi/2\}$ and $\xi=100$. For $\theta_n=\pi/2$ $\alpha^z=0$.}
        \label{fig:sigma12-z-xi100}
    \end{center}
  \end{figure}
such that Eq. \,(\ref{alphabeta}) greatly simplifies to read
\begin{equation}
    -i \Sigma_{12}^{\mu}(P,\theta_n,\xi) = (A-B)\, \alpha^{\mu}(P,\theta_n,\xi)\text{\,,}\label{sigma12}
\end{equation}
where
\begin{equation}
    A-B= {g^2 C_F \over 8 \pi^2} N(\xi) \left[\int_0^\infty dk \,
    k \, \left(f^{\rm iso}_B(k)+f^{\rm iso}_F(k)\right) \right]=\frac{1}{4}m_q^2 N(\xi)\; ,
\end{equation}
assuming equal quark and antiquark distributions.
We did not present the analogous explicit calculation of $\Sigma_{21}$, but find 
for it the same result as for $\Sigma_{12}$ with $A$ and $B$ interchanged.
We also verified that $\Sigma_{12}$ and $\Sigma_{21}$ fulfill the general relation
\begin{equation}
    \Sigma_{21}-\Sigma_{12}=2i\,\text{Im}\,\Sigma\text{\,,}
\label{sigrel}
\end{equation}
with the retarded self-energy $\Sigma$ given in Sec. \ref{quarkse}.
Furthermore, since $\Sigma_{21}$ is given by Eq.~(\ref{sigma12}) with $A$ and $B$ interchanged 
it follows within the hard-loop approximation that with 
the form of the anisotropic distribution function assumed here it always holds that
\begin{equation}
    \Sigma_{12}=-\Sigma_{21}\text{\,,}\label{kms}
\end{equation}
which can be seen as a high-temperature limit for the Kubo-Martin-Schwinger relation in equilibrium, but also holds for finite $\xi$ and 
hence for non-equilibrium.
Eqs.\,(\ref{sigrel}) and (\ref{kms}) show that in order to calculate the hard-loop photon production rate from an anisotropic plasma one need only know the retarded self-energy.
We plot the functions $\alpha$ for an anisotropy parameter of $\xi=100$ and different angles $\theta_n$
in Figs. \ref{fig:sigma12-0-xi100}, \ref{fig:sigma12-x-xi100} and \ref{fig:sigma12-z-xi100} to emphasize the strong angular dependence once more.

\section{Conclusions}

In this paper we have extended the exploration of the 
collective modes of an anisotropic quark-gluon plasma by 
studying the quark collective modes. Specifically, we 
derived integral expressions for the quark self-energy for 
arbitrary anisotropy and evaluate these numerically using 
the momentum-space rescaling introduced in previous papers. 
Using direct numerical calculation we found only real 
timelike fermionic modes and no unstable modes. Additionally 
using complex contour integration we have proven 
analytically in the cases (a) when the wave vector of 
the collective mode is parallel to the anisotropy direction 
with arbitrary oblate anisotropy and (b) for all angles of 
propagation in the limit of an infinitely oblate anisotropy 
that there are no fermionic unstable modes.  Finally, we calculated 
the fermion self-energy of an anisotropic plasma in the 
real-time formalism and demonstrated that within the hard-loop 
approximation the high-temperature limit of the Kubo-Martin-Schwinger 
formula, $\Sigma_{12} = - \Sigma_{21}$, 
holds even for the non-equilibrium configuration considered 
here.  This means that it suffices to only have the retarded 
self-energy $\Sigma$ in order to complete a calculation of 
photon production from an anisotropic plasma in the hard-loop 
framework.  This calculation is currently underway 
\cite{BjornMike}.

\section*{Acknowledgments}
We would like to thank Paul Romatschke and Carsten Greiner for
discussions.

\appendix
\section{Small-$\xi$ limit}

In the limit $\xi\rightarrow0$ we can evaluate the quark
self-energy in a power series in the anisotropy parameter $\xi$.
To linear order in $\xi$ we obtain
\bqa \Sigma_0 &=& \Sigma_0^{\rm iso} + {\xi\over4} \left\{
        {z\over k}\left(3\cos2\theta_n+1\right) +
        \Sigma_0^{\rm iso} \left[ \cos2\theta_n+1
        - (3\cos2\theta_n+1)z^2 \right]
         \right\} , \\
{\Sigma_x\over\sin\theta_n} &=&
        \Sigma_s^{\rm iso} + {\xi\over12} \left\{
        {1\over k}\left(5\cos2\theta_n+3\right) +
        3\Sigma_s^{\rm iso} \left[3\cos2\theta_n+3
        - (5\cos2\theta_n+3)z^2 \right]
         \right\} , \\
{\Sigma_z\over\cos\theta_n} &=&
        \Sigma_s^{\rm iso} + {\xi\over12} \left\{
        {1\over k}\left(5\cos2\theta_n-1\right) +
        3\Sigma_s^{\rm iso} \left[3\cos2\theta_n-1
        - (5\cos2\theta_n-1)z^2 \right]
         \right\} ,
\eqa
where
\bqa \Sigma_0^{\rm iso} &=& {m_q^2 \over 2 k} \log{ \omega +
                k \over \omega -k } \; , \\
\Sigma_s^{\rm iso}&=& {m_q^2 \over k}\left( {\omega\over2k}
            \log{ \omega + k \over \omega -k } -1
            \right)\; .
\eqa

\bibliography{fermions}
\end{document}